\documentclass[twocolumn,prd]{revtex4}
\usepackage{amsmath}
\usepackage{mathrsfs,bm,multirow}
\usepackage{longtable,lscape}
\usepackage{txfonts}
\usepackage{amssymb}
\usepackage{indentfirst}
\usepackage{graphicx,,booktabs}
\usepackage{color}
\usepackage{amssymb}
\begin{document}

\title{A novel configuration of gluonic tetraquark state}
\author{Chun-Meng Tang$^{1}$}
\author{Chun-Gui Duan$^{2,3}$}
\email{duancg@hebtu.edu.cn}
\author{Liang Tang$^{1}$}
\email{tangl@hebtu.edu.cn}
\author{Cong-Feng Qiao$^{4}$}
\email{qiaocf@ucas.ac.cn}
\affiliation{$^1$ College of Physics and Hebei Key Laboratory of Photophysics Research and Application,
Hebei Normal University, Shijiazhuang 050024, China \\
$^2$College of Physics and Hebei Advanced Thin Film Laboratory, Hebei Normal University, Shijiazhuang 050024, China\\
$^3$ Hebei Key Laboratory of Physics and Energy Technology, North China Electric Power University, Baoding 071000, China\\
$^4$ School of Physical Sciences, University of Chinese Academy of Sciences,
Yuquan Road 19A, Beijing 100049, China
}

\begin{abstract}

Inspired by the experimental measurement of the charmed hadronic state X(6900), we calculate the mass spectra of tetraquark hybrid states with configuration of \([8_{c}]_{Q\bar{Q}} \otimes [8_{c}]_{G} \otimes [8_{c}]_{Q\bar{Q}}\) in color, by virtue of QCD sum rules. The two feasible types of currents with quantum numbers $J^{PC} = 0^{++}$ and $0^{-+}$ are investigated, in which the contributions from operators up to dimension six are taken into account in operator product expansion (OPE). In the end, we find that, in charm sector, the tetracharm hybrid states with quantum number \(0^{++}\) has a mass of about \(6.98^{+0.16}_{-0.14} \, \text{GeV}\), while \(0^{-+}\) state mass is about \(7.26^{+0.16}_{-0.15} \, \text{GeV}\). The results overlap with the experimental observations, suggesting potential tetracharm hybrid interpretations. In bottom sector, calculation shows that the masses of tetrabottom hybrid states with quantum numbers $0^{++}$ and $0^{-+}$ are \(19.30^{+0.16}_{-0.17} \, \text{GeV}\) and \(19.50^{+0.17}_{-0.17} \, \text{GeV}\), respectively, which are left for future experimental confirmation.
\end{abstract}
%\pacs{...} % PACS numbers no longer required
\maketitle
\section{Introduction}
In the traditional quark model~\cite{Gell-Mann:1964ewy,Zweig:1964ruk}, hadrons are classified into two types: mesons ($q\bar{q}$) and baryons ($qqq$). However, Quantum Chromodynamics (QCD)~\cite{Gross:1973id,Politzer:1973fx,Wilson:1974sk} allows for the existence of exotic hadron states beyond ordinary mesons and baryons, such as multiquark states, hybrid states, and glueballs, etc. The study of these exotic hadrons provides crucial insights into the non-perturbative behavior of QCD. Over the past few decades, significant advancements in high-energy physics experiments have led to the observation of a series of new hadronic states, such as the XYZ states~\cite{Belle:2003nnu, BaBar:2005hhc, Belle:2011aa, BESIII:2013ris,Belle:2013yex} and hidden-charm pentaquark states~\cite{LHCb:2015yax,LHCb:2019kea}. Since 2003, more than 60 new hadronic states have been discovered in high-energy physics experiments, and many more are expected to be found in the future. Exploring the hadronic structure underlying these experimental discoveries has emerged as a central focus in modern hadron physics research.

In 2020, the LHCb Collaboration reported the observation of a narrow structure, X(6900), with a global significance exceeding 5$\sigma$ in the $J/\psi J/\psi$ invariant mass spectrum, as well as a broad structure in the range of 6.2 to 6.8 GeV, and a hint of another structure around 7.2 GeV in the di-$J/\psi$ spectrum~\cite{LHCb:2020bwg}. This marked the first experimental observation of distinct structures in the double-$J/\psi$ mass spectrum, which is regarded as a major breakthrough in the exploration of hadron spectroscopy. Recently, the ATLAS and CMS Collaborations confirmed the existence of X(6900) in the \(J/\psi J/\psi\) invariant mass spectrum with a significance far exceeding 5$\sigma$~\cite{CMS:2023owd, ATLAS:2023bft}. In the same \(J/\psi J/\psi\) channel, ATLAS also identified two additional structures, X(6400) and X(6600), while CMS observed X(6600) and X(7200) under two models, both with and without considering the interference effects between resonances. Meanwhile, the ATLAS Collaboration reported signals of X(6900) and X(7200) in the \(J/\psi \psi(2S)\) decay channel with a significance exceeding 4.7$\sigma$.

In recent years, fully heavy tetraquark states have attracted significant attention and have been extensively studied across various theoretical frameworks~\cite{Anwar:2017toa, Bedolla:2019zwg, Lloyd:2003yc, Barnea:2006sd, Debastiani:2017msn, Wu:2016vtq, Wang:2019rdo, Liu:2019zuc, Faustov:2020qfm, Lu:2020cns, Heupel:2012ua, Weng:2020jao, Berezhnoy:2011xn, Karliner:2016zzc, Berezhnoy:2011xy, Feng:2020riv, Zhang:2020hoh, Karliner:2020dta, Wang:2020wrp, Giron:2020wpx, Chao:2020dml, Maiani:2020pur, Richard:2020hdw, Zhu:2020xni, Guo:2020pvt, Maciula:2020wri, Zhu:2020snb, Eichmann:2020oqt, Gong:2020bmg, Becchi:2020uvq, Dong:2020nwy, Chen:2016jxd, Wang:2017jtz, Chen:2018cqz, Wang:2018poa, Zhang:2020xtb, Wang:2020dlo, Albuquerque:2020hio, Chen:2020xwe, Wang:2021mma, Wan:2020fsk, Agaev:2023wua, Agaev:2023gaq, Agaev:2023ruu, Agaev:2023rpj, Agaev:2023ara, Jin:2020jfc, Zhao:2020nwy, Wang:2020ols, Wang:2021kfv, Liu:2021rtn, Mutuk:2021hmi, Dong:2021lkh, Wang:2022xja, Wang:2022jmb, Gong:2022hgd, Yu:2022lak, Wang:2021wjd, Liu:2020tqy, Yang:2021hrb, Yang:2020wkh, Tang:2024zvf}. For instance, the authors of Ref.~\cite{Yang:2021hrb} conducted a systematic investigation of these states using a non-relativistic quark model grounded in lattice quantum chromodynamics. They addressed the four-body problem by applying the Gaussian expansion method and complex scaling technique, which revealed the presence of several narrow resonances across all examined tetraquark systems, each exhibiting strong decay widths of less than 30 MeV. In another study, Ref.~\cite{Bedolla:2019zwg} employed a relativized diquark model Hamiltonian to calculate the masses of ground state tetraquarks with quantum numbers $J^{PC}=0^{++}$ and fully heavy baryons across various configurations, producing a comprehensive spectrum for all combinations of heavy quark flavors. Furthermore, the potential for molecular states in the $D^{(*)}\Xi_{cc}^{(*)}$ and $\bar{\Xi}_{cc}^{(*)}\Xi_{cc}^{(*)}$ systems was explored using heavy antiquark-diquark symmetry (HADS). This analysis predicted several meson-baryon and baryon-baryon molecules and suggested a potential connection between the states X(7200) and X(3872), as discussed in Ref.~\cite{Liu:2020tqy}.

In addition, there are also quite a few papers that investigate fully heavy tetraquark states using QCD sum rules. In Ref.~\cite{Wang:2022xja}, the mass spectrum of the ground state and the first two excited states of fully-charm tetraquarks was studied using QCD sum rules and Regge trajectories. By incorporating experimental data, a self-consistent analysis was conducted, and the X(6600), X(6900), and X(7300) were tentatively assigned to tetraquark states with $J^{PC}=0^{++}$ or $1^{+-}$. Additionally, using the inverse Laplace transform sum rule, the masses of doubly-hidden scalar charmonium and bottomonium molecular states were calculated, yielding results consistent with the LHCb experimental observation range of 6.2-6.9 GeV~\cite{Albuquerque:2020hio}.  Moreover, the mass spectra of S-wave and P-wave fully-charm $c\bar{c}c\bar{c}$ and fully-bottom $b\bar{b}b\bar{b}$ tetraquark states in the $8_{[Q\bar{Q}]}\otimes 8_{[Q\bar{Q}]}$ color configuration have been systematically calculated using the moment QCD sum rule method~\cite{Wang:2021mma}, where the predicted masses for the fully-charm $c\bar{c}c\bar{c}$ tetraquark states are approximately 6.3-6.5 GeV for the S-wave channels and 7.0-7.2 GeV for the P-wave channels. Recently, Wan and Qiao proposed a nature hybrid interpretation for the structure of X(6900), i.e. in $[\bar{3}_{c}]_{cc}\otimes [8_{c}]_{G}\otimes [3_{c}]_{\bar{c}\bar{c}}$configuration with $J^{PC}=0^{++}$; by using QCD sum rules, their results showed that the observed X(6900) could be a gluonic tetracharm state, and some other structures may exist, e.g., one around 7.2 GeV in the tetracharm hybrid configuration and with $J^{PC}=0^{-+}$~\cite{Wan:2020fsk}.

In our previous study, we systematically calculated the fully-heavy tetraquark states with various quantum numbers using QCD sum rules for the $[8_{c}]_{Q\bar{Q}}\otimes[8_{c}]_{Q\bar{Q}}$ type currents~\cite{Yang:2020wkh, Tang:2024zvf}. In this work, we propose a new hybrid configuration: tetraquark hybrid state, i.e. $[8_{c}]_{[Q\bar{Q}]} \otimes [8_{c}]_{[G]} \otimes [8_{c}]_{[Q\bar{Q}]}$.
The selection of this specific color configuration (e.g., \([8_{c}]_{Q\bar{Q}} \otimes [8_{c}]_{G} \otimes [8_{c}]_{Q\bar{Q}}\) is rooted in the SU(3) color symmetry constraints. To form a color-singlet hybrid state, the combined color representation of the quark-antiquark pairs and the valence gluon must satisfy: \((([8_{c}]_{Q\bar{Q}} \otimes [8_{c}]_{Q\bar{Q}}) \otimes [8_{c}]_{G}\to 1_{c}\oplus \cdots)\). Only specific combinations (e.g., octet-octet-octet) can achieve an overall color singlet without triviality.

The inclusion of a valence gluon (g) fundamentally differentiates our configuration from the pure tetraquark system or the molecular state. On the one hand, the gluonic component introduces additional dynamical energy via non-Abelian interactions, leading to higher masses compared to pure $qq\bar{q}\bar{q}$ states, consistent with the established theoretical expectation for gluon mass contributions (e.g. $Q\bar{Q}g$ vs $Q\bar{Q}$)~\cite{Qiao:2010zh, Chen:2013zia, Chen:2013eha, Wang:2024hvp}. On the other hand, the valence gluon in our proposed
tetraquark-gluon hybrid introduces distinct decay dynamics compared to pure tetraquark
systems. This gluon-mediated modification alters selection rules and phase-space constraints, potentially suppressing or enhancing specific decay channels. Such decay pattern
anomalies, if observed experimentally, would serve as direct diagnostic evidence for the hybrid nature of these states --- a critical advantage over conventional tetraquark interpretations.
Although lattice QCD and effective field theory studies of this specific configuration are currently absent, experimental anomalies strongly hint at its plausibility. For example, the proliferation of unclassified exotic states (e.g., $P_{c}$ states~\cite{LHCb:2015yax, LHCb:2019kea}, $X(6600)$, $X(6900)$, $X(7200)/X(7300)$~\cite{LHCb:2020bwg, CMS:2023owd, ATLAS:2023bft}, and so on) suggests the existence of unexplored hadronic configurations beyond conventional paradigms.

We perform the calculation of the spectrum of this structure within the framework of the QCD sum rules (QCDSR)~\cite{Shifman:1978bx}. Unlike phenomenological models, QCDSR is a theoretical framework based on QCD, which systematically incorporates non-perturbative effects and offers unique advantages in studying hadron properties involving non-perturbative QCD. It has achieved significant success in hadron spectroscopy research~\cite{Shifman:1978bx, Albuquerque:2013ija, Qiao:2010zh, Chen:2013zia, Chen:2013eha, Wang:2024hvp, Wang:2013vex,Govaerts:1984hc,Reinders:1984sr, P.Col, Narison:1989aq, Tang:2021zti,Qiao:2014vva,Qiao:2015iea,Tang:2019nwv, Wan:2019ake, Wan:2020oxt,Wan:2021vny,Wan:2022xkx,Zhang:2022obn,Wan:2022uie, Wan:2023epq,Wan:2024dmi,Tang:2024zvf,Li:2024ctd,Zhao:2023imq,Yin:2021cbb, Yang:2020wkh,Wan:2024cpc}. To establish QCDSR, the initial step involves constructing suitable interpolating currents that correspond to the hadrons of interest. These currents are then used to create a two-point correlation function, which can be expressed in two ways: the QCD representation and the phenomenological representation. By setting these two representations equal to each other, QCDSR is formally established, enabling the extraction of the hadron's mass.

The remainder of this paper is structured as follows: Section II presents the primary formalism following the introduction. Section III details the numerical analysis and its results comprehensively. Finally, the concluding section discusses and summarizes the key findings.
\section{Formalism}
The QCD sum rules calculations rely on the correlator formed by two hadronic currents~\cite{Shifman:1978bx, Shifman:1978by, Reinders:1984sr, Narison:1989aq, Colangelo:2000dp}. For scalar hadronic currents, the two-point correlation function is expressed as:
\begin{eqnarray}
\Pi(q)&=&i\int d^{4}xe^{iq\cdot x}\langle 0|T\{j(x),j^{\dagger}(0)\}|0 \rangle.
\end{eqnarray}

The constructions of the tetracharm hybrid state with quantum numbers \( J^{PC}=0^{++}, 0^{-+} \) are described as follows:
\begin{eqnarray}
j^{0^{++}}_{A}\!\!&=&\!\!g_{s}f^{abc}\left[ \bar{Q}^{j}\gamma^{\mu}(t^{a})_{jk}Q^{k} \right]G^{b}_{\mu\nu} \left[ \bar{Q}^{m}\gamma^{\nu}(t^{c})_{mn}Q^{n} \right],\label{current-1}\\
j^{0^{++}}_{B}\!\!&=&\!\!g_{s}f^{abc}\left[ \bar{Q}^{j}\gamma^{\mu}\gamma_{5}(t^{a})_{jk}Q^{k} \right]G^{b}_{\mu\nu} \left[ \bar{Q}^{m}\gamma^{\nu}\gamma_{5}(t^{c})_{mn}Q^{n} \right],\label{current-2}\\
j^{0^{-+}}_{A}\!\!&=&\!\!g_{s}f^{abc}\left[ \bar{Q}^{j}\gamma^{\mu}(t^{a})_{jk}Q^{k} \right]\tilde{G}^{b}_{\mu\nu} \left[ \bar{Q}^{m}\gamma^{\nu}(t^{c})_{mn}Q^{n} \right],\label{current-3}\\
j^{0^{-+}}_{B}\!\!&=&\!\!g_{s}f^{abc}\left[ \bar{Q}^{j}\gamma^{\mu}\gamma_{5}(t^{a})_{jk}Q^{k} \right]\tilde{G}^{b}_{\mu\nu} \left[ \bar{Q}^{m}\gamma^{\nu}\gamma_{5}(t^{c})_{mn}Q^{n} \right].\label{current-4}
\end{eqnarray}
Here, $g_s$ denotes the strong coupling constant, $j, k, m, n$ represent color indices, taking values 1, 2, 3, $a$ stands for color indices, taking values from 1 to 8, and $\mu, \nu$ indicate Lorentz indices. $t^a = \frac{\lambda^a}{2}$, where $\lambda^a$ are the Gell-Mann matrices, and $ t^a$ are the generators. The heavy quark field $Q$ stands for either a charm quark or a bottom quark, while $G^b_{\mu \nu}$ signifies the gluon field strength tensor, and $\tilde{G}^b_{\mu \nu} = \frac{1}{2} \epsilon_{\mu \nu \alpha \beta} G^{b, \alpha \beta}$ denotes the dual gluon field strength tensor. It should be noted that the interpolating currents under discussion consist of two color-octet $ \bar{Q}Q$ components along with one gluon, which are analyzed via QCD sum rules for the first time. The adopted interpolating currents for \( J^{PC} = 0^{++} \) and \( 0^{--} \) states adhere to the following configuration criteria, which require that (i) color-octet quark-antiquark pairs form S- or P-wave configurations, with their corresponding color-singlet-state quarkonium states being \( J/\psi \) (S-wave) or \( \chi_{c1} \) (P-wave), respectively, and (ii) the interpolating currents be constructed by coupling them to \( G^a_{\mu\nu} \) or \( \tilde{G}^a_{\mu\nu} \), so as to ensure that these currents match the quantum numbers.

The principle of quark-hadron duality serves as the fundamental assumption for applying the QCD sum rule approach. Therefore, on the one hand, the correlation function $\Pi(q^2)$ can be calculated at the quark-gluon level using the Operator Product Expansion (OPE); on the other hand, it can be expressed at the hadron level, where the hadrons coupling constant and mass are introduced. To calculate the spectral density on the OPE side, the ``full" propagators $S^Q_{jk}(p)$ of a heavy quark (Q = c or b) are utilized:
\begin{eqnarray}
S^Q_{j k}(p) \! \! & = & \! \! \frac{i \delta_{j k}(p\!\!\!\slash + m_Q)}{p^2 - m_Q^2}- \frac{i}{4}g_{s} \frac{t^a_{j k} G^a_{\alpha\beta} }{(p^2 - m_Q^2)^2} [\sigma^{\alpha \beta}
(p\!\!\!\slash + m_Q)\nonumber\\
&+& (p\!\!\!\slash + m_Q) \sigma^{\alpha \beta}]-
\frac{i}{4}g_{s}^{2}(t^{a}t^{b})_{jk} G^{a}_{\alpha\beta}G^{b}_{\mu\nu}\frac{(p\!\!\!\slash + m_Q)}{(p^2 - m_Q^2)^5}\nonumber\\
&\times&(f^{\alpha\beta\mu\nu}+f^{\alpha\mu\beta\nu}+f^{\alpha\mu\nu\beta}) (p\!\!\!\slash + m_Q)\nonumber\\
&+& \frac{i \delta_{j k}}{48} \bigg\{ \frac{(p\!\!\!\slash +
m_Q) [p\!\!\!\slash (p^2 - 3 m_Q^2) + 2 m_Q (2 p^2 - m_Q^2)] }{(p^2 - m_Q^2)^6}\nonumber\\
&\times& (p\!\!\!\slash + m_Q)\bigg\} \langle g_s^3 G^3 \rangle \; ,
\end{eqnarray}
the subscripts \(j\) and \(k\) denote the color indices of the heavy quark. Here, we define \(f^{\alpha\beta\mu\nu} \equiv \gamma^\alpha(p\!\!\!/+m_Q)\gamma^\beta(p\!\!\!/+m_Q)\gamma^\mu(p\!\!\!/+m_Q) \gamma^\nu\). For more detailed information regarding the discussed propagator, readers are encouraged to refer to references~\cite{Albuquerque:2013ija, Wang:2013vex, Reinders:1984sr}.

Thus, based on the dispersion relation, the correlation function $\Pi(q^2)$ at the quark-gluon level can be derived:
\begin{eqnarray}\label{Pi-OPE}
  \Pi^{\text{OPE}}(q^2) = \int_{(4m_{Q})^2}^\infty ds \frac{\rho^{\text{OPE}}(s)}{s - q^2},
\end{eqnarray}
where $\rho^{\text{OPE}}(s) = \text{Im} [\Pi^{\text{OPE}}(s)]/\pi$ and
\begin{eqnarray}\label{rho-OPE}
  \rho^{\text{OPE}}(s) &=& \rho^{\text{pert}}(s) + \rho^{\langle GG \rangle}(s)+ \rho^{\langle GGG \rangle}(s).
\end{eqnarray}
By using the Borel transformation to Eq.~\eqref{Pi-OPE}, we have
\begin{eqnarray}\label{Pi-MB}
  \Pi^{\text{OPE}}(M_B^2) = \int_{(4m_{Q})^2}^\infty ds \rho^{\text{OPE}}(s) e^{-s/M_B^2}.
\end{eqnarray}

\begin{figure}[ht]
  \centering
  \includegraphics[width=8cm]{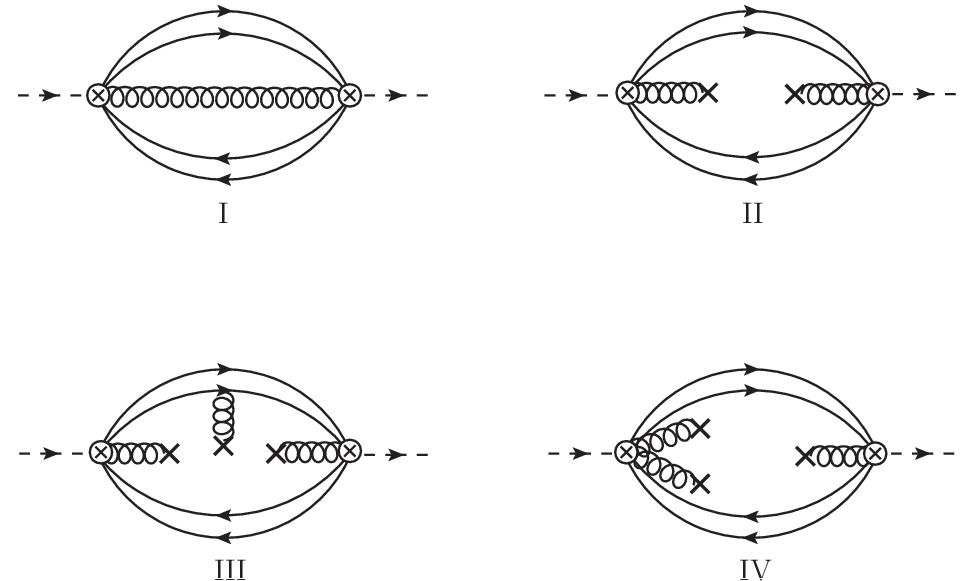}
  \caption{The representative Feynman diagrams for a tetracharm hybrid state, where the permutation diagrams are implicitly accounted for, are shown. Diagrams I, II, III, and IV correspond to the contributions from the perturbative term, two-gluon condensate, and three-gluon condensate, respectively.}
\label{Feyn-Diag}
\end{figure}

The typical leading order Feynman diagrams contributing to Eq.~\eqref{Pi-MB} for the tetracharm hybrid state are shown in Fig.~\ref{Feyn-Diag}. Diagram I represents the contribution from the perturbative part, Diagram II illustrates the two-gluon condensation, and Diagrams III and IV depict the three-gluon condensation. The full expression of the spectral density $\rho^{\text{OPE}}(s)$ is lengthy and will not be shown here.

From a phenomenological perspective, by isolating the ground state contribution of the tetracharm hybrid state, the correlation function \(\Pi(q^{2})\) can be expressed as a dispersion integral over the hadronic spectrum,
\begin{eqnarray}\label{Pi-hadron}
  \Pi(q^2) = \frac{(\lambda_X)^2}{(M_X)^2 - q^2} + \frac{1}{\pi} \int_{s_0}^\infty ds \frac{\rho^h(s)}{s - q^2},
\end{eqnarray}
where $M_X$ is the mass of the  the tetracharm hybrid state, and $\rho^{h}(s)$ is the spectral density that includes contributions from higher excited states and continuum states. $s_0$ denotes the threshold for the higher excited and continuum states. Based on the method detailed and explained in references~\cite{Reinders:1984sr, Colangelo:2000dp}, when calculating the phenomenological aspects of the QCD sum rules, a series of intermediate states must be inserted between the two tetracharm hybrid state interpolating currents. This summation encompasses all possible hadronic states that can be generated by the tetracharm hybrid state currents. The coupling constant \(\lambda_X\) is defined as follows:
\begin{eqnarray}
\langle 0|j^{0^{++}}(0)|X\rangle &=& \lambda_X,
\end{eqnarray}
where $X$ is the lowest lying tetracharm hybrid state. By applying the Borel transformation to the phenomenological side (as described in Eq.~\eqref{Pi-hadron}), and aligning it with Eq.~\eqref{Pi-MB}, we obtain the main equation
\begin{eqnarray}\label{main-equation}
(\lambda_{X})^{2}\exp\left(-\frac{M_{X}^{2}}{M_{B}^{2}}\right)=\int_{(4 m_Q)^2}^{s_{0}} ds \rho^{\text{OPE}}(s) e^{-s/M_B^2}.
\end{eqnarray}
Finally, by differentiating Eq.~\eqref{main-equation} with respect to \(\frac{1}{M_{B}^{2}}\), we derive the mass of the tetracharm hybrid state.
\begin{eqnarray}\label{main-function}
  M_X(s_0, M_B^2) &=& \sqrt{-\frac{L_1(s_0, M_B^2)}{L_0(s_0, M_B^2)}}.
\end{eqnarray}
The moments \(L_{0}\) and \(L_{1}\) are defined as:
\begin{eqnarray}\label{OPE-function}
  L_0(s_0, M_B^2) &=& \int_{(4 m_Q)^2}^{s_{0}} ds \rho^{\text{OPE}}(s) e^{-s/M_B^2}, \label{L0} \\
  L_1(s_0, M_B^2) &=& \frac{\partial}{\partial (M_B^2)^{-1}} L_0(s_0, M_B^2).
\end{eqnarray}

\section{Numerical analysis}
The QCD sum rule expressions involve several input parameters, including the condensates and quark masses. These values have been determined in the literature~\cite{Shifman:1978bx, Shifman:1978by,Reinders:1984sr, Narison:1989aq,Colangelo:2000dp}. For the numerical analysis, we take \(m_c (m_c) = \overline{m}_c= (1.27 \pm 0.03) \; \text{GeV}\) and \(m_b (m_b) = \overline{m}_b= (4.18 \pm 0.03) \; \text{GeV}\), \(\langle g_{s}^{2}G^{2}\rangle=0.88 \; \text{GeV}^{4}\), \(\langle g_{s}^{3}G^{3}\rangle=0.045 \; \text{GeV}^{6}\) where the $\overline{m}_c, \overline{m}_b$ are ``running masses'' of charm and bottom quarks in $\overline{MS}$ scheme.

Additionally, the QCD sum rules introduce two extra parameters, \(M_{B}^{2}\) and \(s_{0}\), corresponding to the Borel parameter and the threshold parameter, which must be determined following the standard procedures. Given a specific \(s_{0}\) value, the Borel parameter \(M_{B}^{2}\) is determined based on two fundamental conditions~\cite{Shifman:1978bx, Shifman:1978by, Reinders:1984sr, Colangelo:2000dp}. To accurately extract the ground state properties of the tetracharm hybrid state, it is first necessary to ensure that the contribution from the continuum spectrum is secondary to the pole contribution (PC)~\cite{Colangelo:2000dp,Matheus:2006xi}. This condition can be expressed through a specific formula:
\begin{eqnarray}
  R_{i}^{\text{PC}}(s_0, M_B^2) = \frac{L_0(s_0, M_B^2)}{L_0(\infty, M_B^2)} \; , \label{RatioPC}
\end{eqnarray}
where the subscript \(i\) represents different currents. Under this condition, the contributions from higher excited states and continuum states are significantly suppressed. This requirement can establish an upper threshold \((M_{B}^{2})_{max}\) for \(M_{B}^{2}\). The second condition concerns the convergence of the OPE. Specifically, it requires comparing the individual contributions with the total magnitude on the OPE side and selecting a reliable region for \( M_B^2 \) to maintain convergence, which will yield a lower bound \((M_{B}^{2})_{min}\) for \( M_B^2 \). In practice, the degree of convergence can be expressed as the ratio of condensates to the total, as follows:
\begin{eqnarray}
  R_{i}^{\text{cond}}(s_0, M_B^2) = \frac{L_0^{\text{dim}}(s_0, M_B^2)}{L_0(s_0, M_B^2)}\, .
\end{eqnarray}
The superscript ``dim'' indicates the dimension of the relevant condensate in the OPE, as described in Eq.~\eqref{L0}. Thus, for a given \(s_0\), the appropriate Borel window, i.e., the range between \((M_{B}^{2})_{min}\) and \((M_{B}^{2})_{max}\), has been determined. In this work, to verify whether the OPE convergence criterion is satisfied, we first ensure that the contribution from the highest-order condensate \(\langle G^{3}\rangle\) does not exceed 10\% of the total OPE contribution.

Certainly, it is necessary to impose a constraint on the value of $s_{0}$, specifically requiring that the mass of the tetracharm hybrid state \(M_{X}\) exhibits minimal dependence on the parameter \(s_{0}\). We employed a method similar to that described in references~\cite{Finazzo:2011he,Qiao:2013raa,Qiao:2013dda} to analyze and accurately determine the appropriate continuum threshold \(s_{0}\). Note that the relationship between \(s_{0}\) and the ground state mass is given by \(\sqrt{s_{0}} \approx (M_{X} + \delta)\) GeV, where \(\delta\) ranges from 0.40 to 0.80 GeV. Therefore, in numerical evaluations, exploring different values of \(\sqrt{s_0}\) is crucial to meeting this criterion. Among these values, we need to select the one that produces the optimal window for the Borel parameter \(M_B^2\). In other words, within the optimal window, the mass of the tetraquark hybrid \(M_X\) is, to some extent, independent of the Borel parameter \(M_B^2\). Thus, the value of \( \sqrt{s_0} \) corresponding to the optimal mass curve will be taken as the central value. In practical applications, we can vary \( \sqrt{s_0} \) by 0.2 GeV in numerical calculations, which sets the upper and lower bounds for \( \sqrt{s_0} \) and helps determine its uncertainty.
\begin{figure}[hbpt]
\begin{center}
\includegraphics[width=6.5cm]{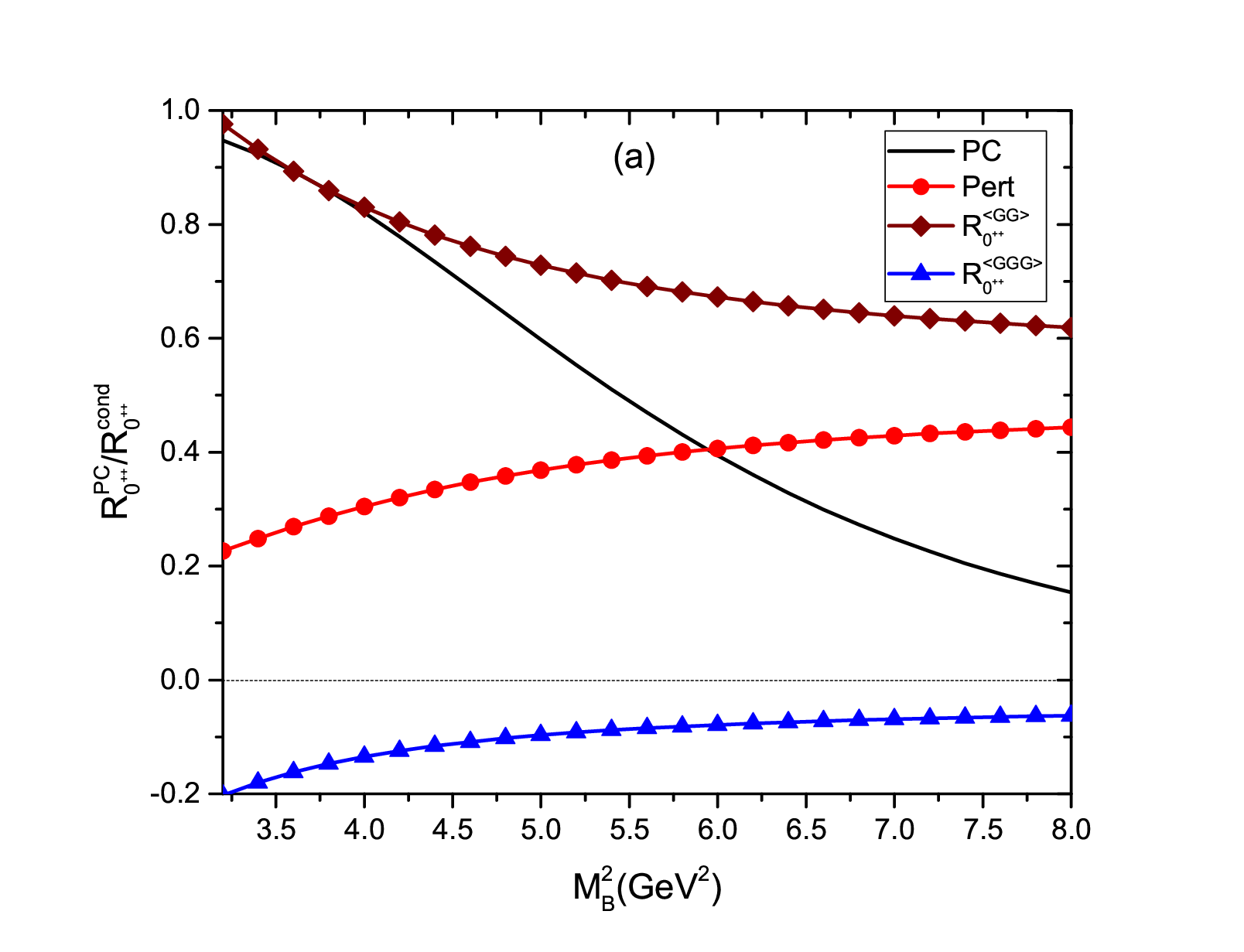}
\includegraphics[width=6.5cm]{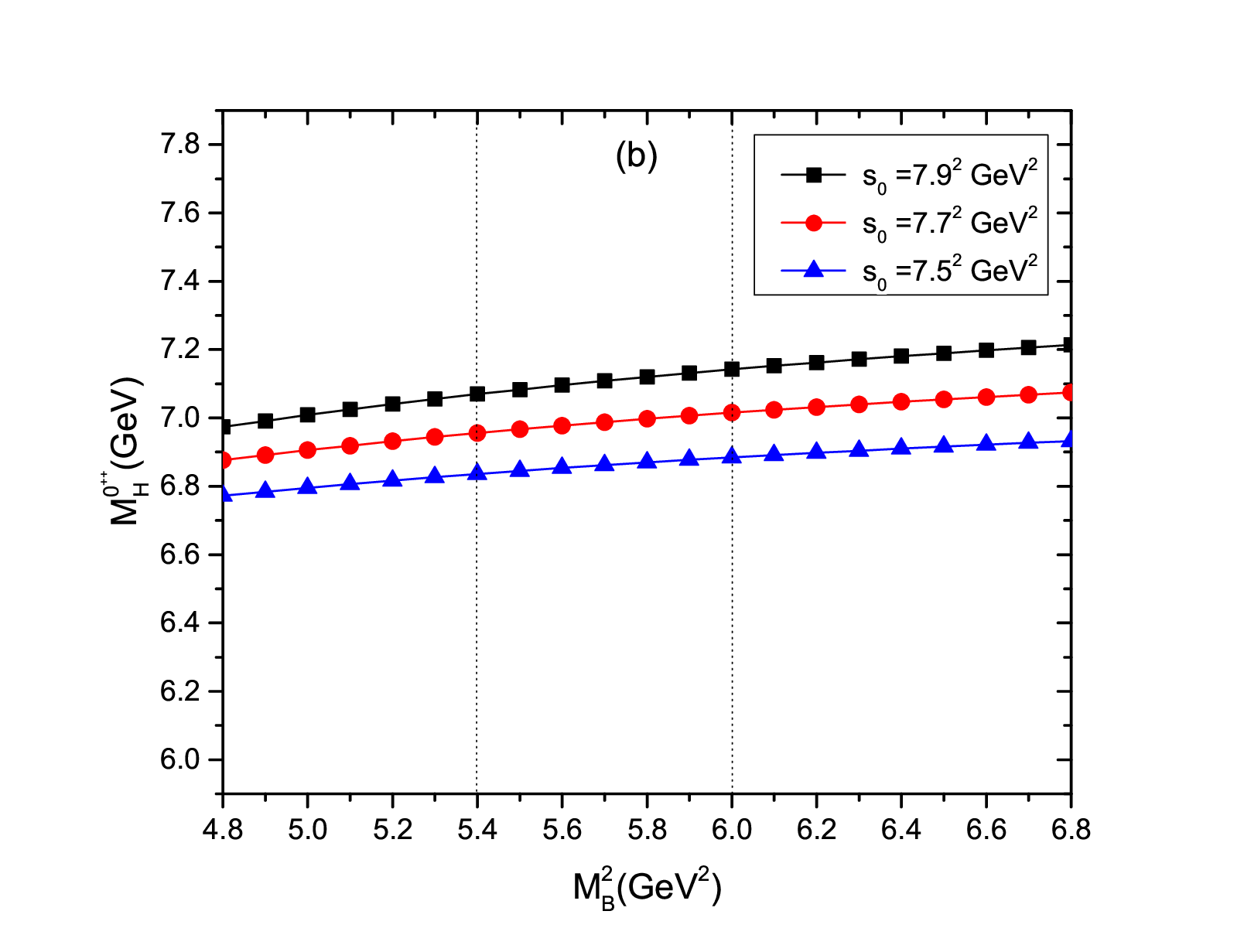}
\caption{Figures of the current \(j_{A}^{0^{++}}\). (a) The pole contribution ratio \(R_{0^{++}}^{\text{PC}}\) and the OPE convergence ratio \(R_{0^{++}}^{\text{cond}}\) as functions of the Borel parameter \(M_B^2\) for the central value of \(s_0\); (b) the mass \(M_{X}^{0^{++}}\) varies with \(M_B^2\), with the selected \(s_0\) values being \(s_0= 7.50^2\, \text{GeV}^2\), \(7.70^2\, \text{GeV}^2\), and \(7.90^2 \, \text{GeV}^2\) from bottom to top. Two vertical lines indicate the upper and lower limits of the effective Borel window with \(s_0\) as the central value. }
\label{fig2}
\end{center}
\end{figure}

\begin{figure}[hbpt]
\begin{center}
\includegraphics[width=6.5cm]{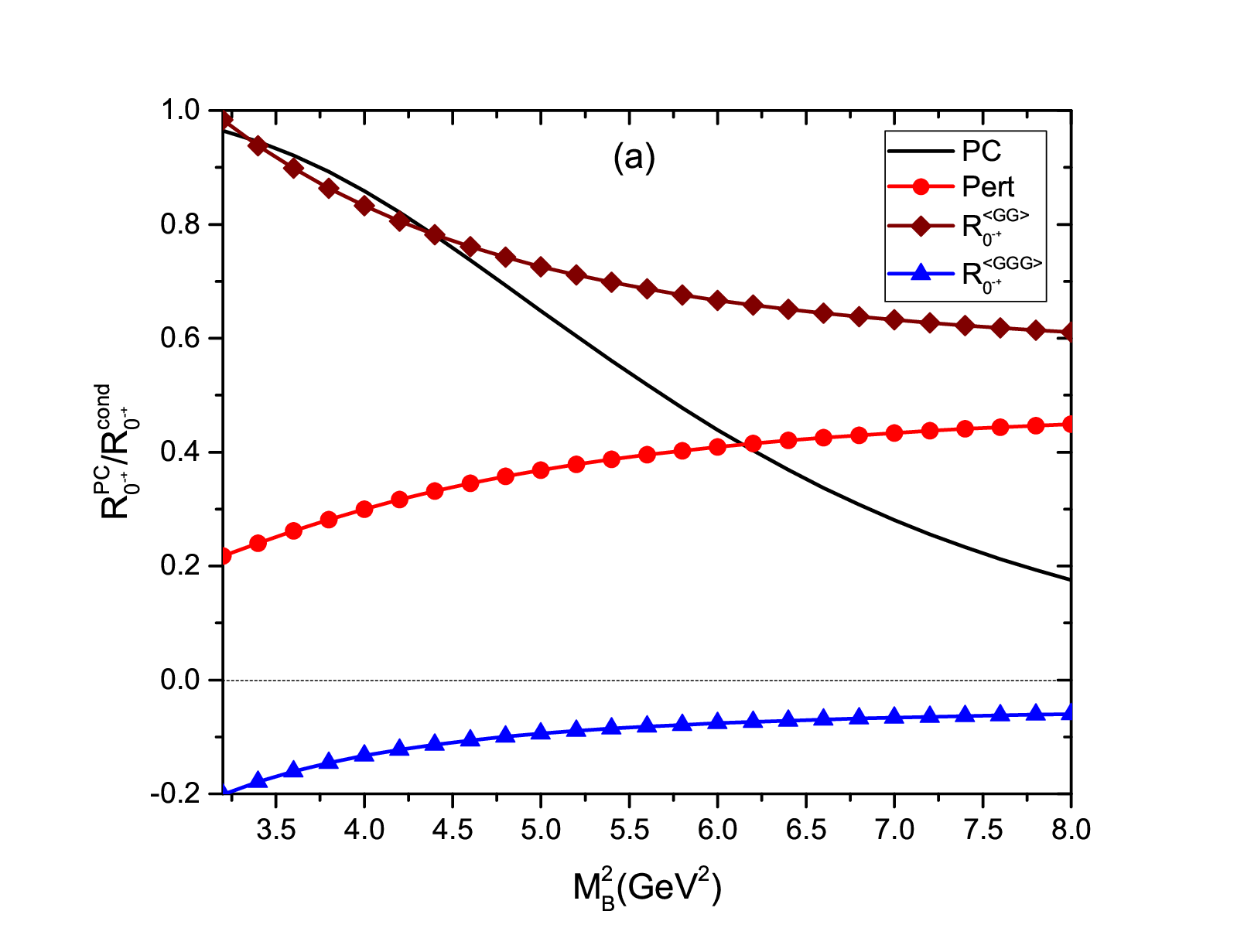}
\includegraphics[width=6.5cm]{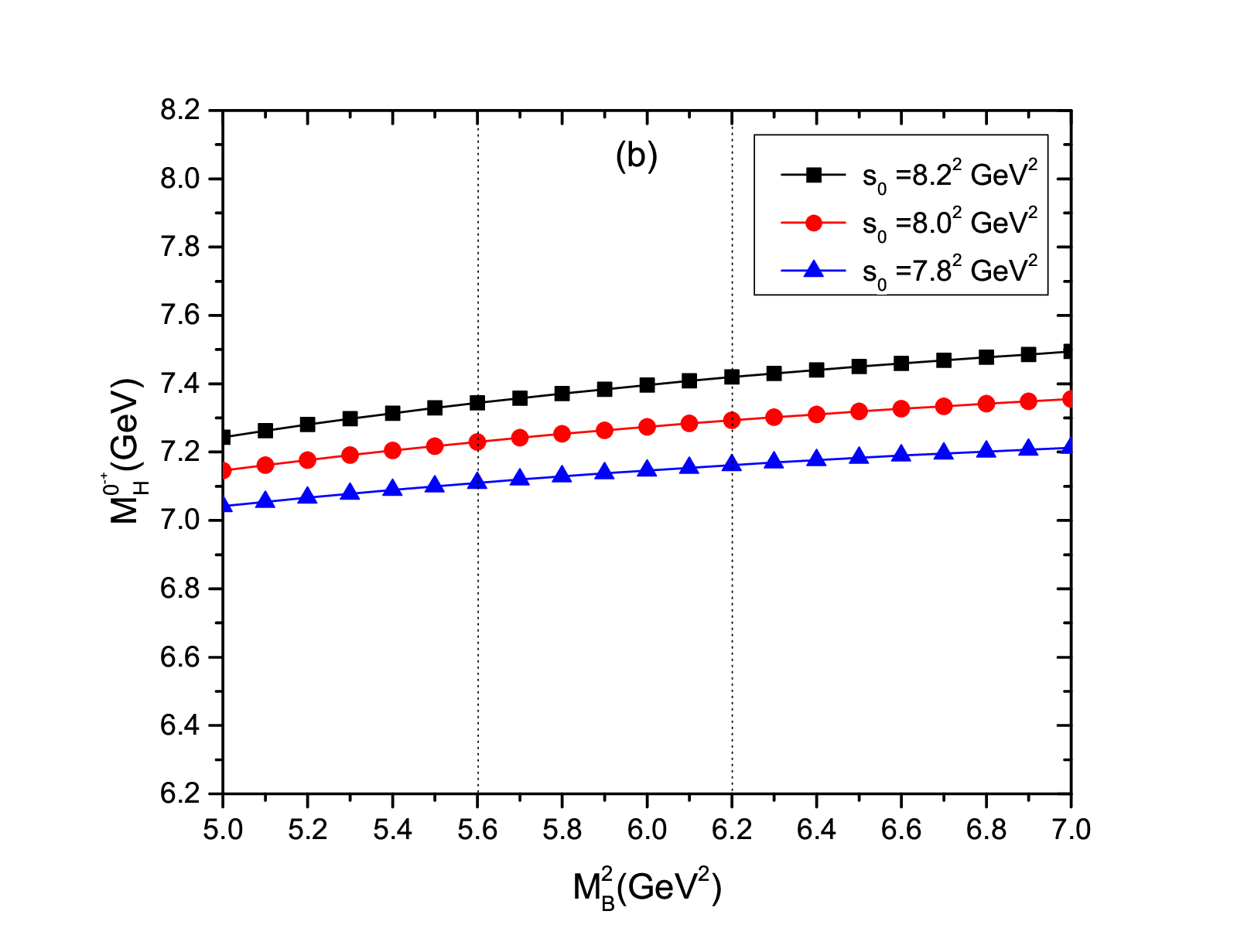}
\caption{The same caption as in Fig.~\ref{fig2}, but for the current \(j_{B}^{0^{-+}}\).}
\label{fig3}
\end{center}
\end{figure}
For the tetracharm hybrid state represented by the current \(j_{A}^{0^{++}}(x)\), characterized by \(J^{PC} = 0^{++}\), we present in Fig.~\ref{fig2}(a) the variation of two ratios, \(R^{\text{PC}}_{0^{++},A}\) and \(R^{\text{cond}}_{0^{++},A}\), as functions of the Borel parameter \(M_{B}^{2}\), with a selected \(s_{0} = 7.70^2\, \text{GeV}^2\). In Fig.~\ref{fig2}(b), we depict the mass relationship with \(M_{B}^{2}\) for different \(s_{0}\) values, also including two lines indicating the upper and lower limits of the effective Borel window under the central \(s_{0}\) value. The selection of $M_{B}^{2}$ and $s_{0}$ complies with the three criteria elaborated earlier. For case $j_{B}^{0^{-+}}(x)$, a similar numerical analysis is conducted, and specific results are presented in Fig.~\ref{fig3}. Notably, the three-gluon condensate contributions for currents $j^{0^{++}}_{A}$ and $j^{0^{-+}}_{B}$ remain below 10\% within each Borel window, satisfying the above-mentioned OPE truncation criterion established, and validating the truncation scheme~\cite{Qiao:2010zh, Chen:2013zia, Chen:2013eha, Wang:2024hvp}.

\begin{table}[htb]
\begin{center}
\begin{tabular}{|c|c|c|c|c|c|c|c|c}\hline\hline
 $J^{PC}$& $M_B^2\,(\rm{GeV}^2)$  & $\sqrt{s_{0}}\,(\rm{GeV})$ & PC & $M_{X}$ $(\rm{GeV})$ & $\lambda_{X}$ $(\rm{10^{-1}GeV^{5}} )$\\ \hline
 $j_{A}^{0^{++}}$ & $5.40\!-\!6.00$ & $7.70 \!\pm\! 0.20$  & $(50\!-\!40)\%$ & $6.98_{-0.14}^{+0.16}$  & $0.99^{+0.08}_{-0.09}$  \\ \hline
 $j_{B}^{0^{-+}}$ & $5.60\!-\!6.20$ & $8.00\!\pm\! 0.20$  & $(50\!-\!40)\%$ & $7.26_{-0.15}^{+0.16}$  & $0.86^{+0.07}_{-0.08}$  \\ \hline
 \hline
\end{tabular}
\end{center}
\caption{The ranges of Borel parameter \(M_{B}^{2}\), the threshold parameter \(s_{0}\), the pole contributions (PC), masses, and the ground state pole residues for currents \(j^{0^{++}}_{A}\) and \(j^{0^{-+}}_{B}\). }
\label{tab1}
\end{table}

For currents \eqref{current-2} and \eqref{current-3}, we could not obtain a positive spectral density function \(\rho_{B}^{0^{++}}\) and \(\rho_{A}^{0^{-+}}\), indicating that the current structures in Eqs.~\eqref{current-2} and \eqref{current-3} do not support the respective tetracharm hybrid states. Therefore, the article only presents numerical curves for the currents \(j^{0^{++}}_{A}\) and \(j^{0^{-+}}_{B}\). The resulting windows of Borel parameter $M_{B}^{2}$, continuum thresholds $s_{0}$, pole contributions, masses, and pole residues are succinctly listed in Table \ref{tab1}. As seen in the table, the dominance of the pole contribution in the phenomenological analysis is well fulfilled within the appropriate Borel window. Within this window, the mass curves display a stable plateau. The central values are consistent with the most stable results concerning \(M_{B}^{2}\), while the error ranges arise from variations in the condensates, quark masses, and the parameters \(s_{0}\) and \(M_{B}^{2}\).

We provide Table.\ref{tab2} to compare our mass predictions and experimental observations.

\begin{table}[htb]
\begin{center}
\begin{tabular}{|c|c|c|c|c|}\hline\hline
  \multicolumn{2}{|c|}{This work}& LHCb~\cite{LHCb:2020bwg} &  CMS~\cite{CMS:2023owd} & ATLAS~\cite{ATLAS:2023bft}  \\ \hline
  $j^{0^{++}}_{A}$ & $6980^{+160}_{-140}$ &$6905\pm11$ &$6927\pm9$ &$6870\pm30$  \\ \hline
  $j^{0^{-+}}_{B}$& $7260^{+160}_{-150}$ &--- & $7287\pm19$  & $7220\pm30$   \\ \hline
 \hline
\end{tabular}
\end{center}
\caption{A comparison between the theoretically predicted masses (in unit of MeV) in this study and the results reported by the LHCb, CMS, and ATLAS experimental collaborations.}
\label{tab2}
\end{table}

Similarly, by replacing the charm quark in theoretical calculations and numerical analysis with the bottom quark, we can predict the masses of tetrabottom hybrid states to be \(19.30^{+0.16}_{-0.17}\) and \(19.50^{+0.17}_{-0.17}\), respectively.

\section{Conclusions}
In this study, we propose a novel hadronic structure composed of two color-octet \(\bar{Q}Q\) components and a color-octet gluon (\([8_{c}]_{Q\bar{Q}} \otimes [8_{c}]_{G} \otimes [8_{c}]_{Q\bar{Q}}\)). We investigate the tetracharm hybrid currents with quantum numbers \(0^{++}\) and \(0^{-+}\). The results reveal that the mass of the tetracharm hybrid state corresponding to the \(0^{++}\) current is \(6.98^{+0.16}_{-0.14}\) GeV, while the mass associated with the \(0^{-+}\) current is \(7.26^{+0.16}_{-0.15}\) GeV. Our analysis suggests that the newly observed states X(6900) and X(7200) correspond to the proposed hadronic structure with $0^{++}$ and $0^{-+}$, respectively, or, at least they may possess significant tetracharm hybrid components.

Table II provides a systematic comparison between our QCD sum rule calculations for the $cc\bar{c}\bar{c}G$ hybrid masses and experimentally measured masses of X(6900) and X(7200) reported by the LHCb, ATLAS, and CMS collaborations. Both theoretical uncertainties and experimental errors are quantified in the table. Crucially, conventional theoretical frameworks (e.g., potential models, lattice QCD) have exclusively addressed tetraquark systems lacking explicit valence gluons --- a fundamental distinction from our tetracharm-gluon hybrid configuration. To date, no rigorous predictions for tetracharm-gluon hybrids corresponding to our proposed
$cc\bar{c}\bar{c}G$ (color-octet-octet-octet) states have been proposed in the literature, underscoring the novelty of this work.

Furthermore, we extended our analysis to the bottom sector and obtained the masses of the corresponding tetrabottom hybrid states with quantum numbers \(0^{++}\) and \(0^{-+}\) as \(19.30^{+0.16}_{-0.17}\) GeV and \(19.50^{+0.17}_{-0.17}\) GeV, respectively. These states may be observed in future experimental studies.

\acknowledgments

This work is supported by the National Natural Science Foundation of China (NSFC) under Grant No. 11975090, 12475087 and 12235008; the Natural Science Foundation of Hebei Province under Grant No. A2023205038 and S$\&$T Program of Hebei (Grant No. 22567617H).

\begin{onecolumngrid}
\appendix
\newpage
\begin{center}
% DO NOT EDIT HERE. Instead edit macro in main.tex to keep
  \textbf{\large\normalfont\bfseries\boldmath Appendix} \\
\vspace{0.05in}

\end{center}

The currents $j^{0^{++}}_{A}$ and $j^{0^{-+}}_{B}$ are represented in the spectral density $\rho^{\text{OPE}}(s)$ as follows:
\begin{eqnarray}
\rho_{0^{++},\,A}^{\text{pert}}(s) &=& \frac{g_{s}^{2}}{2^{12}\times  5\pi^{8}}\int_{x_{i}}^{x_{f}} dx\int_{y_{i}}^{y_{f}} dy\int_{z_{i}}^{z_{f}}dz \left\{ F_{xyz}^{6}(1-x-y-z)xyz\right\} \nonumber\\
&-&\frac{3g_{s}^{2}}{2^{12}\times 5\pi^{8}}\int_{x_{i}}^{x_{f}} dx\int_{y_{i}}^{y_{f}} dy\int_{z_{i}}^{z_{f}}dz \int_{\omega_{i}}^{\omega_{f}}d\omega\left\{ F_{xyz\omega}^{5}m_{Q}^{2}(xy+z\omega)\right\},
\end{eqnarray}

\begin{eqnarray}
\rho_{0^{++},\,A}^{\langle GG\rangle}(s) &=& \frac{\langle g_{s}^{2}GG\rangle}{2^{11}\times\pi^{6}} \int_{x_{i}}^{x_{f}} dx\int_{y_{i}}^{y_{f}} dy \int_{z_{i}}^{z_{f}} dz\left\{F_{xyz}^{2}m_{Q}^{2}(6m_{Q}^{2}-(2F_{xyz}-3s)(x(y-z)\right.\nonumber\\
&-&\left.z(y+z-1)))\right\},
\end{eqnarray}

\begin{eqnarray}
\rho_{0^{++},\,A}^{\langle GGG \rangle }(s) &=& \frac{\langle g_{s}^{3}GGG\rangle }{2^{12}\times\pi^{6}} \int_{x_{i}}^{x_{f}} dx\int_{y_{i}}^{y_{f}} dy\int_{z_{i}}^{z_{f}} dz \frac{1}{xyz(x+y+z-1)}\left\{ 3F_{xyz}m_{Q}^{2}(x^{2}(y+z)\right.\nonumber\\
&+&\left.x(y^{2}+y(2z-1)+(z-1)z)+yz(y+z-1))(2m_{Q}^{2}- (F_{xyz}-s)(x(y-z)\right.\nonumber\\
&-&\left.z(y+z-1))) - 6(4F_{xyz}^{3}xyz(x+y+z-1)+3F_{xyz}^{2}(m_{Q}^{2}(x(y-z)\right.\nonumber\\
&-&\left.z(y+z-1))- 6sxyz(x+y+z-1))-2F_{xyz}(m_{Q}^{4}+3m_{Q}^{2}s(x(y-z)\right.\nonumber\\
&-&\left.z(y+z-1))-6s^{2}xyz(x+y+z-1)) - s(-m_{Q}^{4}+m_{Q}^{2}s(x(z-y)\right.\nonumber\\
&+&\left.z(y+z-1))+s^{2}xyz(x+y+z-1))) \right\},
\end{eqnarray}

\begin{eqnarray}
\rho_{0^{-+},\,B}^{\text{pert}}(s) &=& \frac{g_{s}^{2}}{2^{12}\times  5\pi^{8}}\int_{x_{i}}^{x_{f}} dx\int_{y_{i}}^{y_{f}} dy\int_{z_{i}}^{z_{f}}dz \left\{ F_{xyz}^{6}(1-x-y-z)xyz\right\} \nonumber\\
&+&\frac{3g_{s}^{2}}{2^{12}\times 5\pi^{8}}\int_{x_{i}}^{x_{f}} dx\int_{y_{i}}^{y_{f}} dy\int_{z_{i}}^{z_{f}}dz \int_{\omega_{i}}^{\omega_{f}}d\omega\left\{ F_{xyz\omega}^{5}m_{Q}^{2}(xy+z\omega)\right\},
\end{eqnarray}

\begin{eqnarray}
\rho_{0^{-+},\,B}^{\langle GG\rangle}(s) &=& \frac{\langle g_{s}^{2}GG\rangle}{2^{11}\times\pi^{6}} \int_{x_{i}}^{x_{f}} dx\int_{y_{i}}^{y_{f}} dy \int_{z_{i}}^{z_{f}} dz\left\{-F_{xyz}^{2}m_{Q}^{2}(6m_{Q}^{2}+(2F_{xyz}-3s)(x(y-z)\right.\nonumber\\
&-&\left.z(y+z-1)))\right\},
\end{eqnarray}

\begin{eqnarray}
\rho_{0^{-+},\,B}^{\langle GGG \rangle }(s) &=& \frac{\langle g_{s}^{3}GGG\rangle }{2^{12}\times\pi^{6}} \int_{x_{i}}^{x_{f}} dx\int_{y_{i}}^{y_{f}} dy\int_{z_{i}}^{z_{f}} dz -\frac{1}{xyz(x+y+z-1)}\left\{ 3F_{xyz}m_{Q}^{2}(x^{2}(y+z)\right.\nonumber\\
&+&\left.x(y^{2}+y(2z-1)+(z-1)z)+yz(y+z-1))(2m_{Q}^{2}+ (F_{xyz}-s)(x(y-z)\right.\nonumber\\
&-&\left.z(y+z-1))) - 6(4F_{xyz}^{3}xyz(x+y+z-1)-18F_{xyz}^{2}sxyz(x+y+z-1)\right.\nonumber\\
&+&\left.2F_{xyz}(m_{Q}^{4}+ 6s^{2}xyz(x+y+z-1))-s(m_{Q}^{4}+s^{2}xyz(x+y+z-1))) \right\},
\end{eqnarray}

Here, we use the following definition:
\begin{eqnarray}
  F_{xyz} &=& m_{Q}^{2}f_{xyz} - s, \\
  F_{xyz\omega} &=& \bigg(\frac{1}{x}+\frac{1}{y}+\frac{1}{z}+\frac{1}{w}  \bigg)m_Q^2-s\;,\\
  x_{f/i}&=&\bigg[\bigg( 1-\frac{8m_Q^2}{s} \bigg) \pm \sqrt{\bigg( 1-\frac{8m_Q^2}{s} \bigg)^2-\frac{4m_Q^2}{s}}\bigg] \bigg/2\;,\\
y_{f/i}&=&\bigg[ 1+2 x +\frac{3 s x^2}{m_Q^2-s x} \pm \sqrt{\frac{[m_Q^2+s x (x-1)][(8x+1)m_Q^2+s x(x-1)]}{(m_Q^2-s x)^2}}  \bigg] \bigg/2\;,\\
z_{f/i}&=&\bigg[(1-x-y)\pm \sqrt{\frac{(x+y-1)[m_Q^2(x+y-(x -y)^2)+s x y(x+y-1)]}{s x y-(x+y) m_Q^2}} \bigg]\bigg/2\;,\\
w_{-}&=&\frac{x y z m_Q^2}{s x y z -(x y +y z + x z)m_Q^2}\;,w_{+}=1-x-y-z\;.
\end{eqnarray}

\end{onecolumngrid}

\end{document}